# Framing the Fundamental Theorem of Calculus Through Physics-Based Quantities

Suzanne White Brahmia[1*] and Patrick Thompson[2]

[1*]Department of Physics, University of Washington, Box 351560, Seattle, 98195-1560, WA, USA.
[2]School of Mathematical and Statistical Sciences, Arizona State University, Box 871804, Tempe, 85287-1804, AZ, USA.

*Corresponding author(s). E-mail(s): brahmia@uw.edu;
Contributing authors: pat@pat-thompson.net;

**Abstract**

There is a substantial curricular overlap between calculus and physics, yet introductory physics students often struggle to connect the two. We introduce a quantity-based framing of the Fundamental Theorem of Calculus (FTC) to help unify learning across both disciplines. We propose a consistent approach to teaching definite integrals, including shared vocabulary and symbolism, to help students recognize how concepts like change, rate, and accumulation show up in both calculus and physics. We argue that the typical interpretation of the FTC in calculus, focusing on antiderivatives in closed form, doesn't align well with how physicists use or conceptualize integration. We advocate for an additional focus on Riemann sums and the underlying ideas of change, rate, products, and accumulation, which are fundamental in both fields. This approach can help students build a deeper, more coherent understanding of both mathematics and physics quantity. By aligning learning objectives across the disciplines, we argue that students can develop a stronger understanding of foundational mathematical principles.

---

[1]SWB would like to acknowledge the funding from the United States National Science Foundation DUE-2417104 that has helped support this work.

# 1 Introduction

Physics is the science of change, quantified by an abundance of different physical quantities. Calculus, in turn, offers the formal tools to describe how these quantities vary and relate, providing a structure within which new physical quantities can emerge. The two disciplines–physics and calculus–are deeply intertwined. Recognizing this, most STEM curricula require students, particularly those in physics and engineering, to complete calculus and calculus-based introductory physics courses early in their academic careers. These courses are intended to prepare students to reason quantitatively and to apply calculus meaningfully in physical contexts. However, research shows that many students perceive a disconnect between doing mathematics and reasoning mathematically within physics contexts. This perceived divide limits their capacity to engage in meaningful quantitative reasoning. For instance, in a study by Taylor and Loverude (2023), a student described the disconnect plainly: "I have math and physics on different days, so I forget about math when I go to physics, I forget about physics when I go to math." This comment highlights a broader issue: although students encounter overlapping concepts in math and physics, they often fail to incorporate them.

This disconnect is especially problematic given the central role that variable quantities play in physics. While introductory physics courses introduce over a hundred physical quantities, instruction relies on a narrow set of familiar functions to describe their variation – many of these quantities share common covariational structures. Helping students recognize these patterns could enable them to apply calculus-based reasoning more fluently in physics.

This paper argues for a shared instructional goal across calculus and physics: that students understand why they use calculus–not just how to perform its procedures. Drawing on the framework of *proceptual* understanding of Gray and Tall (1994), we advocate that symbolic operations in calculus should evoke quantitative meaning, and vice versa. In other words, students should not only be *pro*cedurally fluent but *conceptual*ly grounded in their use of calculus to model changing quantities.

To support this goal, we present a quantity-centered framing of the Fundamental Theorem of Calculus (FTC) rooted in physics modeling. Our approach emphasizes that quantities–carefully defined and contextually meaningful–form the conceptual bridge between mathematics and physics. Physical modeling enables prediction and explanation by connecting measurable quantities through mathematical relationships. In this view, each calculus operation serves a purpose grounded in physical reasoning. For example, consider the quantity of flux. Broadly, flux measures how much of something passes through a surface, a concept involving both rate and amount. Depending on the context, flux may be defined as a rate per unit area (e.g., particle flux) or as a measure of a field through a surface (e.g., electric or magnetic flux). Despite differing mathematical treatments, the core idea remains consistent: flux quantifies the movement of something through a real or imaginary surface. In electricity and magnetism, integrating flux over a surface relates to the quantity of a source, such as charge or current. In thermodynamics, particle flux connects to thermal properties of a system. Across domains, the physical world motivates the mathematics; calculus becomes necessary, not optional.

The framing we propose is designed to support instructors in organizing their teaching around essential physics quantities and their mathematical representations. We envision this structure guiding the design of instructional activities, shaping classroom discussions, and fostering interdisciplinary connections between calculus and physics. More specifically, we provide a quantity-based framing of the FTC for instructors aiming to create learning environments in which students:

- Spend meaningful time exploring the foundational ideas of amount, change, rate of change, interval, and accumulation in contexts that matter to them. Leveraging students' prior knowledge of physics can help them make sense of calculus. Naming and describing quantities, not just symbolizing them, supports engagement with and deeper understanding of physics.
- Prioritize conceptual understanding over procedural speed (Thompson, 1994). Instruction should connect mathematical procedures to their conceptual roots. This includes explicit discussions that deepen students' understanding of how mathematical expressions adapt across different contexts, with attention to the meanings of symbols and how variable values change.

The physical world provides a natural context in which calculus becomes intellectually necessary (Harel, 2008). This necessity creates opportunities for deeper learning. Our goal is to provide a structure through which calculus instructors can meaningfully integrate physics-based quantities, thus supporting student learning of the FTC and promoting transfer across disciplines.

We align our work with recommendations from Ely and Jones (2023):

> "Reasoning with definite integrals is a key skill for calculus students to develop as part of their curriculum, and the ability to interpret integrals in the context of modeling with quantities is critical to the learning of calculus."

Despite recent efforts to include modeling in mathematics courses, many physics courses do not expect students to engage in genuine calculus reasoning. Physics textbooks often avoid situations where variable quantities are combined with other variable quantities, largely because students are typically ill-equipped mathematically to reason with them. As a result, many relationships are simplified to constant-rate approximations to keep the physics storyline manageable (Loverude, 2025).

Von Korff and Rebello (2012) provide evidence of the instructional challenge that quantitative reasoning with variable quantities poses in physics. In a case study, they conducted a series of teaching interviews with a student, for a total of 14 hours throughout the term, to characterize and support her understanding of definite integrals in mechanics contexts, guided by insights from mathematics education research (Zandieh, 2000). While their approach showed promise in helping the student make connections between calculus and physics, as an intervention it is not feasible to scale up. Much of their effort focused on helping the student develop foundational ideas that could have been introduced earlier in calculus instruction – rate, change, and accumulation.

This paper responds to the need for calculus students to develop a deeper understanding of the foundational ideas of rate, change, and accumulation. Avoiding calculus in physics instruction shortchanges students, and the disconnect between the disciplines undermines both. We argue that coordinated instructional efforts between calculus and physics are not only necessary but achievable, and our physics quantity-based framing of the FTC presents one step in that direction.

In the sections that follow, we outline our framing and its research foundations, and make recommendations for its uptake as well as future research directions. Specifically, in §2 we review relevant research on student learning and knowledge construction, highlighting both key resources and difficulties students bring from calculus into physics, as well as a current quantities-focused framing of the FTC from the mathematics research literature. In §3, we build on existing quantities-focused FTC research and extend it into the realm of physics, illustrating how the FTC functions as a profound knowledge structure that can support students' reasoning in physics. In §4, we discuss research findings which reveal current learning obstacles that our framing can help instructors navigate. In §5, we present a physics quantity-based framing of the FTC that can inform both physics and calculus instruction. This framing meets the objectives of both addressing the obstacles from the prior section, and bridging disciplinary divides in support of more coherent learning trajectories. Lastly, in §7, we outline directions for future research and development and discuss existing instructional materials that can serve as a foundation.

## 2 Background

### 2.1 Conceptual foundations of the FTC

In the context of a calculus course both historically and conventionally, integrals are introduced as representing areas under curves in a Cartesian coordinate system. This geometric interpretation is a powerful abstraction rooted in a basic quantitative principle: if a quantity $Q$ changes at a constant rate $q(x)$ over an interval of length $\Delta x$, then the change in $Q$, denoted $\Delta Q$, is given by $q(x)\Delta x$. This product can be represented graphically as the area of a rectangle with height $q(x)$ and width $\Delta x$.

When the rate of change $q(x)$ varies over the interval $[x, x+\Delta x]$, we can approximate it as constant over that interval to obtain an estimate for $\Delta Q$. As $\Delta x$ becomes smaller, the approximation improves, and summing these over the interval $[a,b]$ yields an increasingly accurate estimate of the total change in $Q$. In the limit as $\Delta x \to 0$, the sum approaches the exact change in $Q$, and thus, the area under the graph of $f(x)$ over $[a,b]$ comes to represent the total change in a quantity whose rate of change is given by $f(x)$. This connection underlies the conventional association of definite integrals with the area under a curve.

However, it is important to emphasize that this geometric interpretation is specific to Cartesian coordinates. It does not hold in other coordinate systems, such as polar or semi-logarithmic systems, where the relationship between area and accumulated change is not preserved in the same way.

Over time, area bounded by curves have become the customary meaning of integrals. The fact that the integral $\int_a^b f(x)dx$ represents a value of a quantity not having values $f(x)$ or $x$ is often lost in presentation and discussion. Students learn that an integral is an area. Also lost is the fact that in $\int_a^b f(x)dx$, every value of $f(x)$ is a *rate of change of an accumulating quantity* with respect to a quantity having value $x$.

The Fundamental Theorem of Calculus (FTC) is typically stated in two parts, as in §5.3, (Briggs et al., 2011):

Let f be a continuous function on an interval $[a, b]$. Let the function F be defined as $F(x) = \int_a^x f(t)dt$. Then

1. $F'(x) = f(x)$          ($F$ is defined as the antiderivative of $f$)

2. $\int_a^b f(x)dx = F(b) - F(a)$          (evaluate the values of $F$ at boundaries)

It is important to note that $x$ is the independent variable in the definition of $F$. There are natural interpretations of $x$ and $t$ when $F$ is interpreted as an accumulation function. The value of $x$, the upper limit of integration, varies when modeling any quantity that accumulates. As for $t$, for any value of $x$, that is for any specific accumulation, the value of $t$ varies from $a$ to $x$, which gives us a specific value for accumulation for a specific value of $x$. The meaning of $F$ then is the net accumulation in the quantity being modeled for any interval of accumulation determined by $a$ and $x$.

In the first equation above, we see that the meaning we must give $f$ in $\int_a^x f(t)dt$ is that values of $f$ are values of the accumulating quantity's rate of change with respect to the quantity whose value is $x$. This is not to say that the original meaning of $f(x)$ must be a rate of change. Instead it says that the value of the accumulating quantity's rate of change with respect to $x$ is identical to the value of $f(x)$. If the accumulating quantity has unit $U_F$ and the independent quantity has unit $U_x$, then $f(x)$, as a rate of change of accumulation, will have the unit $U_F$ per $U_x$.

The customary significance of the first equation above in the FTC comes from the (usually previously established) fact that any two antiderivatives of $f$ differ at most by a constant. So, if you can find a function $G$ defined in closed form whose derivative is $f$, then you can calculate the value of $F(b)$, or $\int_a^b f(x)dx$, by calculating $G(b) - G(a)$. This standard FTC interpretation allows for efficiently hand-calculating definite integrals whenever a closes-form antiderivative exists, which was invaluable before the age of computers. But it comes at a price. The connections among concepts of change, rate, products and accumulation, which are essential to the mathematical sense that physicists make with physical quantities is, lost.

Consider now, the Fundamental Theorem of Calculus as a relationship between fundamental mathematical quantities[2]. In his foundational work on calculus, Isaac Newton introduced the terms *fluent* and *fluxion* to describe what are now understood as functions and their derivatives, respectively. A fluent denoted a quantity that varies continuously over time, while a fluxion represented its instantaneous rate of change. Newton framed the essential problems of the calculus in two parts: first, to determine the fluxion given a fluent–what is now recognized as the process of differentiation; and second, to recover the fluent from its fluxion–what we now term indefinite integration. For instance, given the *fluent* $x(t) = v_o t$, its *fluxion* for all values of $t$ is $v_o$. While Newton's notation and terminology were initially influential, they were ultimately replaced by the differential and integral notation developed independently by Leibniz.

---

[2] There are two ways we can consider the FTC – in the mathematical context of a standard calculus course and in its quantitative significance.

A productive understanding of the FTC for physics is, in effect, similar to Newton's ideas of change and variation–considering all quantities as flowing or having flowed. When quantities flow, they have a rate of change with respect to some other quantity. This rate of change is the rate at which the quantity accumulates. In our characterization, a productive understanding of situations as embodying the FTC is based in these ways of seeing the world.

- If you understand a quantity as varying, it occurs to you immediately that any value is an amount of accumulation. The quantity built to that amount.
- If you understand two quantities as varying in relation to each other, it occurs to you immediately that each quantity's value varies at some rate of change with respect to the other.
- If you understand two quantities' values as varying at some rate of change with respect to each other, it occurs to you immediately that their values accumulate with respect to each other.

We emphasize that the above dispositions are *pre-symbolic*. They are ways of seeing the world, not ways of interpreting mathematical statements. They provide individuals with a disposition to see situations as modeled appropriately with integrals or derivatives, which embody rates and accumulation. These ways of thinking are themselves dependent on students developing other dispositions earlier in their schooling, regarding *creating quantity* – having a disposition to ask, "What is being measured? How is it measured? What does a particular measure mean?", and *variation* – imagining total variation is an accumulation of small variations[3] .

The FTC relates the rate of change of quantity A with respect to quantity B with the accumulation of quantity A in relation to quantity B by way of summing the product of its rate of change over infinitesimal intervals of change in its independent quantity and the size of those intervals. It tells a rich story of the interplay between quantities as they change.

Figure 1 illustrates several important ideas about the relationship between a function $f$ and its accumulation function $F$. First, although $f$ does not have an elementary (closed-form) antiderivative, it still has a well-defined antiderivative in the sense of the Fundamental Theorem of Calculus:

$$F(x) = \int_a^x f(t)\,dt.$$

Figure 1 demonstrates this relationship by approximating the derivative $F'(x)$ using the difference quotient:

$$r_F(x) = \frac{F(x+h) - F(x)}{h}.$$

On the left, where $h = 1$, the approximation of $F'(x)$ is relatively poor. On the right, with a much smaller step size ($h = 0.001$), the approximation significantly improves. At this resolution, the graphs of $y = f(x)$ and $y = r_F(x)$ are nearly indistinguishable, visually confirming that $F'(x) = f(x)$.

---

[3] It is unfortunate that many students are mystified as to how an area in a coordinate system gives an amount of distance, work, or force (Jones, 2015).

In other words, Figure 1 illustrates that at each point $x$, the value of $f(x)$ corresponds to the instantaneous rate of change of the accumulation function $F$. Figure 1 embodies two additional aspects.

1. The function $F$ defined as $F(x) = \int_a^x f(t)dt$ is treated in this graphing program as a first-class function –it can be evaluated, graphed, composed with other functions, etc. We propose that students actively defining accumulation functions and using those definitions as first-class functions can be a benefit to their conceptions of integrals as mathematical objects that have meaning in the context being modeled.[4]

2. The use of function notation is central to mathematics at levels of precalculus and beyond. However, function notation is used in physics (too) sparingly. This is a potential roadblock for students in relating calculus in mathematics with calculus in physics.

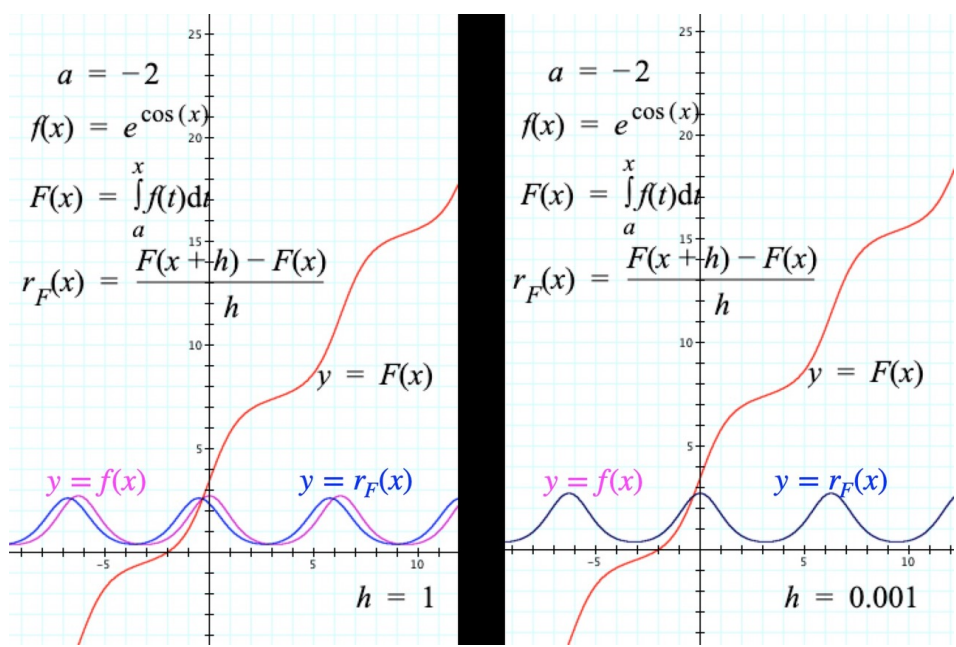

**Fig. 1** An accumulation function $F$ defined in open form and its approximate rate of change function $r_F$ for two values of $h$. Left: $h = 1$; Right: $h = 0.001$. The graphs of $f$ and $r_F$ appear to coincide over sufficiently small intervals–illustrating that $F'(x) = f(x)$. Notice that while $f$ does not have an elementary (closed-form) antiderivative, it does have an antiderivative–namely $F(x) = \int_a^x f(t)dt$

---

[4]This is akin to calls for computational mathematics, wherein the introduction or use of any major mathematical quantity is accompanied by the matter of how to produce a value of it. (CITE Peters-Burton, 2020

To retain a focus on quantification that is essential in students' physics courses, and better aligned with Newton's framing, we advocate a focus of calculus teaching that elevates Riemann-sum reasoning (summing up products of two quantities – a rate of change and an infinitesimal change) and de-emphasizes a focus on finding closed-form antiderivatives to solve a problem. Antiderivatives achieved their prominence historically because they enabled physicists and mathematicians to hand-calculate definite integrals. Today this is a nicety, but computing devices today allow acceptable approximate solutions expressed as summations. Wagner (2018) provides evidence for the cognitive dissonance students experience based on their educational experiences focused on the importance of the antiderivative in solving definite integrals. He contends that Riemann sum-based reasoning doesn't align with solution processes using antiderivatives. The use of computing devices for calculating Riemann sums can produce acceptable approximations to definite integrals. The foundational mathematical quantities of accumulation, change and rate of change are central to enriching the learning of calculus, and its role in the quantification of physics.

## 2.2 Quantification and physics quantities

Quantification is a foundational cognitive process in both mathematics and physics, yet its role is often under emphasized in standard calculus instruction. In physics, conceptual understanding begins with identifying and defining quantities–attributes of physical phenomena that can be measured or calculated–and then establishing meaningful relationships among them. Many of these relationships are multiplicative or proportional in nature, forming the basis for reasoning with rates of change and accumulation about systems in motion.

Quantities such as momentum, force, and density are not introduced arbitrarily. Rather, they emerge through conceptual reasoning grounded in intuitive or experiential understanding. For example, momentum arises from the idea that both mass and velocity contribute to an object's "quantity of motion." The relationship $p = mv$ reflects the intuition that doubling either mass or velocity should double the momentum. Similarly, density is not just the result of dividing mass by volume; it expresses how mass is distributed in space, formalized as $\rho = \frac{m}{V}$. These quantities are constructed, not just computed: they encode relationships that must be interpreted as meaningful, not merely manipulated symbolically.

Quantification also involves the ability to symbolically represent quantities and their relationships. Students must learn to interpret and use letters and symbols not merely as placeholders in equations but as representations of measurable, variable quantities–each with units, and often with direction or sign. Vector quantities, for example, require additional representational fluency. Notational conventions such as vector hats (e.g., $\hat{\imath}, \hat{\jmath}, \hat{k}$), subscripts (e.g., $F_x, v_{\text{initial}}$), and signed scalars (e.g., positive and negative values for direction) are not simply formal embellishments; they convey crucial conceptual information about orientation, reference frames, and interaction. These representational demands complicate quantification but are essential for modeling physical systems accurately.

Further, distinctions between variables, parameters, and constants are critical to understanding how quantities behave across and within physical situations. As Thompson and Carlson (2017a) clarify, *parameters* are quantities treated as fixed within a particular context, though they may vary from situation to situation. In contrast, *variables* change within the context of a single scenario. Confusing these roles can obscure the structure of a mathematical model and hinder students' reasoning in both physics and mathematics. As Philipp (1992) emphasizes, keeping track of how a symbol is being used–whether as a parameter, constant, or variable–is vital for meaningful interpretation of mathematical expressions. We emphasize that the heavy use of symbolizing in physics contexts renders this distinction essential to understanding models.

Research has demonstrated that reasoning grounded in quantification and proportionality supports more robust mathematical understanding. For instance, Ellis (2007) found that students who engaged in *emergent-ratio reasoning*–constructing ratios from relationships between quantities–were more successful in generalizing about linearity and providing valid justifications than those who relied on pattern-based or procedural strategies. Similarly, Moore et al. (2009) reported that attending to students' construction of quantitative relationships within context enabled them to engage more successfully in mathematical modeling. These findings suggest that reasoning about quantities in context strengthens both conceptual mathematical understanding and transfer to new problems.

The absence of this grounding in many calculus classrooms contributes to a persistent disconnect between formal mathematical procedures and physical meaning. In a study by Bajracharya et al. (2023), mathematics majors were asked to make sense of a negative definite integral. The researchers found that invoking a physical context–a stretched spring–helped students understand the meaning of $dx$ as representing a small physical displacement, rather than just a symbolic directive for integration. One student remarked that the context led them to realize that "$dx$" was not simply a variable for use in symbolic manipulation, but "represented something"–a small, measurable quantity. This example illustrates the cognitive power of grounding calculus in physical interpretation, as well as the epistemological divide many students perceive between "pure" mathematics and physical reasoning.

Taken together, these findings argue for a more deliberate integration of quantification into calculus instruction. Helping students develop a flexible and meaningful understanding of how quantities are defined, related, and represented–both symbolically and conceptually–offers a pathway toward deeper mathematical reasoning and greater relevance to scientific contexts.

## 2.3 Symbolic forms and symbolic blending frameworks

Contrary to popular "separate world" models in mathematics education, in which mathematics and the rest of the world occupy separate mental spaces (Blum and Leiß, 2007), mathematics education researchers studying student problem-solving report that students engage in a continuous contextual validation of their mathematization (Czocher, 2016; Sealey, 2014; Borromeo Ferri, 2007). These findings are consistent with other research that shows expert physics modeling as a tight blend of physics and mathematical worlds (Zimmerman et al., 2025), and that physics majors reason productively when they blend these worlds (Schermerhorn and Thompson, 2023; Van den Eynde, 2021). Even introductory physics students are more efficient when they blend these worlds (Kuo et al., 2013). The interplay between quantities and mathematics is inseparable in physics. We approach this work treating the mathematics that is used in physics as a conceptual blend of physics quantities and mathematical objects. The conceptual blending framework is a theory of cognition developed by Fauconnier and Turner in which elements from distinct scenarios are "blended" in a subconscious process (Fauconnier and Turner, 2002).

We consider quantities and mathematical objects to be inseparable, nonetheless they are made up of small pieces of blended knowledge. Cognitive resources are the fine-grained pieces of knowledge that people use to create a thought, and mathematics with quantities is the grain size of thought in physics. Resources are knowledge structures a person draws upon to understand and solve physics problems, including elements like declarative, experiential and procedural knowledge, spatial reasoning, visual imagery etc. that are needed to effectively engage with the problem. The resources students combine when reasoning with definite integrals in physics contexts are not the techniques for solving an integral but consist of smaller components like concepts of rate, change, derivatives, summations, or differentials. Arguably, knowledge-in-pieces is consistent with the Riemann sum-based reasoning, and at best neutral regarding an antiderivative representation.

According to coordination class theory of learning, resources are organized and reorganized over time; as students move through their course curricula, they develop coordinated sets of resources that represents repeated patterns in reasoning (diSessa and Wagner, 2005). Sherin (2001) studied third-semester calculus-based physics students as they collaboratively solved unfamiliar problems, and developed a framework for categorizing *symbolic forms*–compact cognitive structures that integrate procedural and experiential knowledge about quantities and their representations.

Dorko and Speer (2015) applied Sherin's framework to analyze how calculus students reason about area and volume. They developed a "measurement" symbolic form that combines numeric value and unit as one. The authors observed that students who wrote correct units could explain dimensions of planar figures and solids, and connect this knowledge to the shapes' units. In contrast, students who struggled with units also struggled with dimensionality. White Brahmia (2019) extended the work of Dorko and Speer and developed a "quantity" symbolic form central to physics reasoning which establishes sign ($+$ or $-$) as a feature of physics quantity, in addition to numeric value and unit because of the varied and essential information the sign carries about the physical quantity (White Brahmia et al., 2020).

One of Sherin's forms that is particularly relevant in the context of this paper is the "parts-of-a-whole" form, where a whole quantity is made up of several smaller, additive parts, typically indicated by multiple terms added together using plus signs ($+$). Meredith and Marrongelle (2008) investigated how students apply integration in solving electrostatics problems in introductory calculus-based physics, analyzed through Sherin's symbolic forms framework. They found that students have many strong, purely-mathematical resources but often struggle to apply them in physics contexts.

One set of resources that were productive involved Riemann sums; the researchers characterized students notions of summing up small contributions as Sherin's "parts-of-a-whole" form, and observed students used it frequently and productively to guide their work solving definite integrals in this context.

Jones (2013) investigated how experienced calculus students understand and conceptualize integration, focusing on the role of symbolic forms in their reasoning. Through this analysis, Jones highlighted the significance of the "adding up pieces" interpretation within the symbolic forms framework. His findings support the perspective shared by other researchers that emphasizing accumulation and the process of summing infinitesimal contributions can enhance students' conceptual grasp of the integral and improve their ability to apply it flexibly across varied contexts. Oehrtman and Simmons (2023) developed an emergent model of how introductory calculus students construct and interpret definite integrals to model physical quantities. The authors emphasize the importance of quantitative reasoning and a parts-of-a-whole framing in students' understanding of definite integrals.

Many researchers in physics education agree that mathematics and physics are a conceptual blend, based on Fauconnier and Turner's conceptual blending framework (Bing and Redish, 2009; White Brahmia et al., 2021). Schermerhorn and Thompson (2023) introduce a "symbolic blending" theoretical framework, which combines Sherin's symbolic forms and Fauconnier and Turner's conceptual blending frameworks in physics. The symbolic blending model combines the mathematical structure of equations (symbolic forms) with the contextual understanding of physics concepts (conceptual blending). The authors argue for the benefits of a symbolic blending model for disentangling mathematical justification from contextual knowledge in physics, even though students and experts are holding both in mind at all times. Symbolic blending facilitates envisioning a framework in which student resources can be fostered in both calculus and physics courses, with each retaining their own disciplinary learning objectives.

Although mathematics and physics are inseparable for physicists, mathematics courses can play a critical role in strengthening students' understanding of the mathematical objects that underpin physical quantities. That strengthening must happen in the contexts of physical quantities; we propose several in §3, as well as a structure for supporting students' learning of key calculus objects in §5.

## 2.4 FTC: Parts-of-a-whole symbolic form

We've established that quantities can enhance students' understanding of what they are doing when they solve a definite integral, and that a parts-of-a-whole symbolic form is productive in a variety of contexts for students in calculus and in physics. The mathematical abstractions of rate, accumulation, product and change as relationships between physical quantities are so important in physics, that frequently they become new quantities and given their own name, and are connected through the Fundamental Theorem of Calculus (FTC).

In this section we present a formulation of the FTC that focuses on mathematical abstractions and their symbolic representations (Thompson, 1994; Samuels, 2022), that lends itself well to adaptation in physics contexts. Samuels (2022) framework focuses on the quantities of amount, change, rate and accumulation (ACRA). The author applies the ACRA formulation to the evaluation theorem of the FTC:

$$\int_a^b F'(x)dx = F(b) - F(a)$$

**Table 1** The ACRA framework: Mathematical abstractions and the FTC (Samuels, 2022)

| FTC: | $F(b) - F(a)$ | $= \int_a^b dF$ | $= \int_a^b \frac{dF}{dx} dx$ | $= \int_a^b F'(x) dx$ |
|---|---|---|---|---|
| | **Total Change** | **Accum** | **Accumulation** | **Accumulation** |
| | Total change | Infinite sum of dep. variable change | Infinite sum of dep. variable change for each input change × input change | Infinite sum of rate × input change |

Table 1 represents Samuel's ACRA framework. $F(x)$ is the value of the accumulating quantity represented by the dependent variable, and $x$ is the value of a quantity represented by the independent variable. $F(x)$ and $x$ are amounts, $F'(x)$ is a rate of change (RoC) of accumulation with respect to $x$. The right side of the equation represents the **change in $F(x)$**. On the left side is an integral, or an infinite sum, which also represents **an accumulation**. The terms being summed are each the product of the RoC of $F(x)$, and an (infinitesimal) change in $x$. The authors contend that the four essential quantities of calculus are: Amount, Change, Rate, Accumulation. An important emphasis, both in instruction and in the assessment of learning, is that $F'(x)$ is a rate–the RoC of accumulation with respect to variations in $x$. These four quantities (Amount, Change, Rate, and Accumulation) and relationships among them form the ACRA framework.

In physics contexts, an amount would be the measured value, including its units, of a quantified property (Thompson and Carlson, 2017b). A variable is used to represent an amount of something. In the next section, we leverage Samuel's ACRA framework of the FTC to connect these ideas to foundational relationships between quantities in physics through the FTC.

## 3 An FTC framing of physics modeling

Within the first weeks of the electromagnetism course, which is typically taken concurrently with the integral calculus course where the FTC is first mentioned, physics students are expected to learn new abstract physical quantities through definite integrals. The integrals relate electric force and field to their sources in an intricate story of the interplay between these vector quantities. It is assumed that students immediately recognize that they are summing up products and quantifying the accumulation.

As an example of the complexities involved for the introductory physics student, we turn our focus specifically to Gauss's Law (see Fig. 2), for its rich blend of mathematical abstractions – varied symbolizing, multiplicative structures involving both vector and scalar physical quantities, and a vector-valued differential. Gauss's law equates the electric flux (the accumulation on the left-hand side of the equation) to the total amount of electric charge, $Q_{enc}$, which is enclosed by the imaginary shape. The labeled arrows radiating out from the shape represent the vector values of the electric field and area differential, respectively, at several points on the surface. The two sides of the equation are made equal by the inclusion of a physical constant, $\varepsilon_o$. An alternate representation makes the rate-change product more transparent here, consistent with ACRA. The product of $\varepsilon_o$ and $\vec{E}$ is a quantity, $\vec{D}$, known as the *flux density*, or the area RoC of the flux, rendering Gauss's law: $\oint \varepsilon_o \vec{E} d\vec{A} = \oint \vec{D} d\vec{A} = Q_{enc}$.

Gauss's law demonstrates that a solid understanding of the integral as a sum of small products can help students begin understanding what is being said here, despite the heavy symbolizing and other abstractions. If students can rely on mathematics to help guide them here, they can immediately see the integrand is tiny bits of something that result from the dot product of an electric field vector and a small interval area vector. The integral sums up the small bits to find the total flux. All of that reasoning can take place fairly straightforwardly by minimizing the distraction from the complexities of the mathematics. Thereby, students' minds are freed to focus on the physics notion that the net flux is proportional to the charge enclosed by the surface – which is one of the four fundamental ideas that form the basis of classical electromagnetism, represented as one of Maxwell's equations.

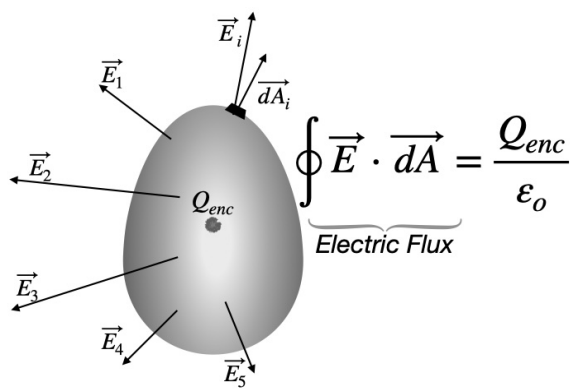

**Fig. 2** Gauss's law relates the electric flux to the amount of charge enclosed in the Gaussian surface. The accumulation on left hand side is the total flux, and it is proportional to the charge enclosed, represented by $Q_{enc}$.

Li and Singh (2017) report that students learning Gauss's law struggle with understanding the principle of superposition (despite the summing up that is represented by the integral) and making a distinction between the electric field (vector quantity, represented by the $\vec{E}$), and the electric flux, the (scalar) accumulation. Student difficulties recognizing the process of summing up small products quantities in physics integrals is common. In Section 4 we present other evidence from research in much less abstract contexts.

Gauss's law is far out of scope as a context for a calculus course, but is shared here as an example of the complexity that calculus takes on in introductory physics which can be simplified if students come prepared to think in terms of the framing characterized in Table 1.

In this section we described connections, as centralized in Table 2, that can be made in the context of the fundamental theorem of calculus (FTC), building on the ACRA framing of Table 1. The top three columns of Table 2 replicate Table 1 and then extend it into the physics domain by delineating examples where the ACRA framing shows up in the very first course in physics. Building on these examples in a calculus course can better equip students to conceptualize the relationships between rates of change, accumulation and other quantities that are central to physics models using examples that are within the zone of proximal development for students and instructors. Supporting students' quantitative reasoning with these contexts can help prepare them to use that reasoning in more complicated ones, like Gauss's law.

A first course in calculus-based introductory physics typically spans kinematics, Newton's laws and the conservation of energy and momentum. Most courses start by developing the ideas foundational to the *kinematics equations* – position and velocity equations as a function of time for constant acceleration motion. Our framing starts there as well.

## 3.1 Kinematic Equations

The first day of physics typically begins with a description of motion in one dimension at a steady speed. The concept of a time RoC first appears in physics with velocity as:

$$x(t) = x_o + v_o t$$

The quantity *acceleration*, $a$, is soon introduced as the time RoC of velocity. By analogy, a kinematics equation for constant acceleration motion in one dimension is introduced.

$$v(t) = v_o + at$$

A non-zero *change* in the velocity vector, $\Delta \mathbf{v} = \mathbf{v}_f - \mathbf{v}_o$, indicates that the system is accelerating and therefore, by Newton's 2nd law, there is an unbalanced force acting on it. $\Delta \mathbf{v}$ is one of the most important quantities that guides thinking in introductory physics; the scalar components of vector quantities are used in one dimension for simplicity, as written above. Note that finding the difference between the initial and final values of the velocity is one way of determining its change, but the FTC provides another way using rate, multiplication and sum to determine an *accumulation* that is generalizable beyond the standard constant-acceleration motion of kinematics.

$$\int_a^b F'(x)dx = F(b) - F(a)$$

Summing up the rate (the acceleration) multiplied by each (infinitesimal) time interval results in an *accumulation* that is $\Delta v$. In the case of constant acceleration, the integral is more machinery than is necessary. But the act of setting up this mathematical machinery in the context of learning about the FTC can set students up to see how rates – even non-constant ones– are used to calculate change. Quantifying rates of change and using them to help determine change is a recurrent pattern in physics (and other STEM) modeling that can be introduced through the use of quantities in calculus instruction.

An example of a changing rate students will encounter in their first week of physics is the uniformly changing velocity described by the kinematic equations. The accumulated effect is a displacement, $\Delta x$ that can be found using the FTC, more generally allowing for any form of variation in the velocity,

$$x(t) = x_o + \int v(t)dt$$

$$\Delta x = \int v(t)dt$$

See Table 2 for a summary of these recurring patterns in the kinematics equations, and the structure for the pattern as outlined in the FTC.

## 3.2 Conservation Laws

Energy and momentum are students' entry points to understanding the conservation laws in physics, with each remaining constant in a closed system. When the system is not closed, the mechanisms of work and impulse quantify changes in energy and momentum, respectively. These conservation laws serve as foundational principles, guiding reasoning across all areas of physics. The FTC offers a symbolic template for representing changes in these quantities, linking accumulated change to rates of change in a mathematically coherent way.

The total energy of a system is changed when an object from outside the system exerts a force on the system while it changes position along the line of the force. The change in the system energy is equal to the accumulated effect of a force acting over a distance. This accumulation is so important, it is a named quantity– *work*. Here, the force both causes and quantifies the rate at which work is done as the position varies.

The total momentum of a system is changed when an object from outside the system exerts a force on the system over *a time interval*. The change in the total momentum is equal to the accumulated effect of a force acting over a time interval. This accumulation is also so important, it is a named quantity– *impulse*. Here, the force both causes and quantifies the rate at which the momentum changes as time varies. See Table 2 for a summary of these recurring patterns in the conservation laws.

**Table 2** Foundational Quantities in Physics and the FTC. Values of $F$ are accumulations over an interval of its argument

| | $F(b) - F(a)$ | $= \int_a^b dF$ | $= \int_a^b \frac{dF}{dx} dx$ | $= \int_a^b F'(x) dx$ |
|---|---|---|---|---|
| **Quantity** | **Change** | **Change** | **Rate $\times$ Change** | **Accumulation** |
| | Total change | Infinite sum of dep. variable change | Infinite sum of dep. variable change for each input change $\times$ input change | Infinite sum of rate $\times$ input change |
| $\Delta v^{1,2}$ | $v(t_2) - v(t_1)$ | $= \int_{t_1}^{t_2} dv$ | $= \int_{t_1}^{t_2} \frac{dv}{dt} dt$ | $= \int_{t_1}^{t_2} a(t) dt$ |
| | Change in velocity | Same as above | Same as above | Infinite sum of acceleration $\times$ time interval |
| displacement[1] | $x(t_2) - x(t_1)$ | $= \int_{t_1}^{t_2} dx$ | $= \int_{t_1}^{t_2} \frac{dx}{dt} dt$ | $= \int_{t_1}^{t_2} v(t) dt$ |
| | Change in position | Same as above | Same as above | Infinite sum of velocity $\times$ time interval |
| work done on system[3] | $U(x_2) - U(x_1)$ | $= \int_{x_1}^{x_2} dU$ | $= \int_{x_1}^{x_2} \frac{dU}{dx} dx$ | $= \int_{x_1}^{x_2} F(x) dx$ |
| | Change in system potential energy | Same as above | Same as above | Infinite sum of force $\times$ displacement |
| impulse[4] | $p(t_2) - p(t_1)$ | $= \int_{t_1}^{t_2} dp$ | $= \int_{t_1}^{t_2} \frac{dp}{dt} dt$ | $= \int_{t_1}^{t_2} F(t) dt$ |
| | Change in system momentum | Same as above | Same as above | Infinite sum of force $\times$ time interval |

We emphasize that the significance, and practical differences, between the independent and the dependent variables here go beyond their positions in the equation. Force causes systems to change resulting in an accumulated change in quantities. The recurring patterns that are reflected in the quantities that make up the FTC provides a way of thinking about the relationships among various quantities, providing a learning opportunity for students to more deeply understand foundational ideas in both mathematics and physics. We argue that Table 2 is just a sample of the many other contexts where this framing appears in the introductory physics sequence.

---

[1] Kinematics
[2] Newton's Laws
[3] Conservation of Energy
[4] Conservation of Momentum

In the next section, we describe research foundations for the quantity-based framing of the FTC that we make in the §5.

# 4 FTC physics framing: Research foundations

The symbolic blending of the FTC quantities in physics contexts presents a significant challenge each time students encounter new topics in physics. At the heart of this difficulty is sensemaking with quantities and operations, and their meaning both in a mathematical sense and a physical one. We'll provide evidence here of obstacles to making mathematical meaning in the context of physically realistic contexts.

Specifically, the evidence will reveal obstacles:

- **Overemphasis on an exact, continuous function as the only correct solution:** Emphasizing closed-form antiderivatives over a Riemann sum interpretation of the FTC reinforces a view of a single immutable model.
- **Confounding limits and infinitesimals**: The differential represents a physical quantity, which cannot disappear. Students' image of an amount approaching zero presents a cognitive barrier for many of them.
- **Symbolizing of quantities can render meaning opaque**: Moving from $f(x)$ to, say $P(V)$, and understanding the meaning of $d$ in $dx$ can be render meaning-making very challenging for students, especially when the symbols represent physical quantities.

## 4.1 Exactness and correctness

Many students complete their study of calculus without being able to interpret the definite integral as a sum, conflating the techniques they've learned to do in pursuit of a continuous, closed form solution with the mathematical meaning of the integral. Jones (2015) argues this is possibly due to the overemphasis on area-under-a-curve and antiderivative techniques and underemphasis on Riemann sum-based reasoning in their instruction, a priority that reinforces a view of exact, continuous functions as the definitive and correct solutions in mathematics. This framing aligns with mathematical values, where exactness, continuity, and formal derivation often define what counts as a correct answer.

However, this perspective can be misleading when applied to physics. In physics, correctness is rooted in empirical evidence: mathematical models are valued not for their formal exactness but for how well they approximate and explain observed data. The disciplinary differences are transparent in a study conducted by Roundy et al. (2015) with faculty in mathematics, physics, and engineering. The subjects were asked to measure a specific derivative $\frac{dx}{dF_x}$ using a device that allowed them to make (and measure) changes in $F_x$ (an interval $\Delta F_x$) and measure the resulting changes in $x$. They report that the physicists and engineers immediately set to task designing a way to measure a derivative by measuring the change in the dependent variable x over intervals of $\Delta F_x$ and calculating the ratio. The mathematicians spent much of the interview making meaning of the symbols used, and eventually set to work collecting data to find a function that they could differentiate symbolically. Through interaction with the interviewer, they did not consider a computed average RoC to be a derivative, regardless of its precision. The physicists and engineers knew that their computation was an approximation, but they also knew how to ensure that it was a good one. We argue that the mathematicians in this context were conflating exactness with correctness. Figure 1 demonstrates a value of $h$ that is only as small as it needs to be in order to meet the need, in this case the need is the screen resolution of "sameness".

An equation-based model in physics is an idealized representation of real-world patterns, not an absolute truth. The epistemology of science holds that if future data reveals a better-fitting model, then that new model is considered more correct. Thus, while mathematics may prioritize the elegance and precision of closed-form solutions, physics treats mathematical functions as provisional tools–approximations subject to revision based on evidence.

## 4.2 Infinitesimals in the zero limit

The results of the Roundy et al. (2015) experiment exemplify an important disciplinary rift, namely the meaning made by "$dx$". Some people mean it as a signal for the variable of integration. Others think of it as a vanishing amount "going to zero", as in $\lim dx \to 0$. Others think of both "$dx$" and "$dy$" as variables related by $dy = f'(x)dx$, in the tradition of Fréchet. Others yet think of "$dx$" as an infinitesimal magnitude as in the tradition of Robinson's non-standard analysis–a number that is greater than 0 and smaller than any positive real number. Finally, others, mainly scientists, think of "$dx$" as meaning an amount of a quantity small enough to produce acceptably accurate results in computations.

We see the Fréchet interpretation (values of $dx$ and $dy$ vary) and the scientific interpretation (small enough to give acceptable approximations) as consistent and mutually supportive, and both being compatible with Robinson's notion of infinitesimal. Fréchet's approach provides a conceptual foundation for linear approximation even at the level of infinitesimal change. Thompson et al. (2019) melded these three meanings into their development of integrals and derivatives without formally stating any one of them, which is also reflected in the Samuels (2022) ACRA framework.

The Zandieh (2000) model of students' understanding of derivatives offers a useful framework for identifying barriers to transferring calculus knowledge to physics contexts. The model conceptualizes understanding in terms of hierarchical "layers." At the most basic level is the ratio layer, where the derivative is understood as a ratio of two finite quantities. The next is the limit layer, which requires students to imagine the denominator of that ratio approaching zero. At the highest level is the function layer, where the derivative is conceived as a function in its own right. Importantly, conceptual understanding–not just procedural fluency–at each layer is necessary to build toward the next. This layered view helps illuminate why many students, even those with strong procedural skills, struggle to apply derivatives meaningfully in physical situations.

Layer one is foundational to a Riemann-sum interpretation for the integral. Regarding layer two, there is ample evidence that students struggle to make meaning of a differential, specifically in the context of it being infinitesimal–not zero, but not different from 0 by any positive real amount. Oehrtman (2009) reports on student reasoning in which there is a collapse in dimension, "...corresponding to the independent variable in the limit ... going to zero, this dimension was ultimately imagined to vanish."

Physics can, and does, accommodate an interval that is slightly greater than the zero. While students can follow this reasoning, the heavy emphasis on $dx$ pointing to a variable of integration or $dx$ being infinitesimally small in their math courses, with no alternative interpretation, leaves them feeling hesitant to engage in this kind of "sloppiness" in physics. Meredith and Marrongelle (2008) found that the notion of a limit going to zero can hinder students' understanding of integrals as sums, especially when they don't understand that $dx$ is never 0 in $\lim_{dx \to 0,\, dx \neq 0}$. The authors argue that students need guided instruction to reinterpret mathematical concepts in physics contexts, where summation of all pieces is a foundational idea. Jones (2013) observed an unproductive resource of adding up the integrand, which supports Meredith and Marrongelle's and Oehrtman's observations of confusion that can set in when obliged to think of differentials uniquely in the context of limits. Nguyen and Rebello (2011) asked students to interpret infinitesimal intervals of area, $\vec{dA}$, in a physics context. Even though they productively used a summing up reasoning to interpret the integral, they felt that $\vec{dA}$ refers to a changing area (process), rather than a small element of area (quantity). The authors contend that helping students to make a distinction between a process and a quantity may require some instruction targeted at the student's physical intuition about infinitesimals. Von Korff and Rebello (2012) state that, in their approach that has shown promise, "the integral can also be constructed by summing an 'infinite' number of infinitesimal products, although a traditional calculus framework would not allow this."

Increasingly, mathematics education researchers argue for the pedagogical value of treating differentials as infinitesimal change in calculus (Ely and Jones, 2023; Ely, 2017; Thompson and Dreyfus, 2016). The authors present evidence that framing change as a small quantity supports students in making meaningful sense of the *mathematics* they are engaging with.

There are also arguments that differentials play a key pedagogical role in understanding calculus. Modeling two continuously covarying quantities–rather than discrete, incremental changes–can better support mathemtics students' conceptual development of limits. For example, Castillo-Garsow et al. (2013) present case studies of high school students reasoning about variation and argue that smooth images of change (differentials) are more powerful than "chunky" ones in contexts of covariation. As they note:

> "Chunky thinking generates chunky conceptions of variation, whereas smooth thinking generates smooth conceptions... A smooth conception involves attending to all states continuously, without privileging unit values that invite counting. In contrast, chunky conceptions always yield countable products, no matter how small the chunk."

We note here that most work in mathematics on student thinking with quantities has been done in the contexts of independent variables that students experience – time, volume of water flowing through a pipe, etc. Mathematics students can readily visualize what appears to them to be a continuous process. But the physical world is not continuous in extremely small scales, which is a realm where a lot of physics takes place. There are units that do privilege a basis for counting and they never go to zero. The notion of $\lim dx \to 0$ is in disagreement with physics in the small. The objectives of physics and of mathematics can be quite different. In physics, the mathematics that students encounter models the physical world. Continuous functions can be used as approximations to the world they describe, not the other way around. The physical world is a chunky place.

We argue here that chunky infinitesimals that are not nearly equal to zero are consistent with the realities of the physical world. Zero-limit infinitesimals may serve an important purpose as a procedural cue to the variable of integration and, operationally, there is nothing wrong with that approach - it helps you efficiently get an answer. But as a method it has strayed far from Newton's notions of why you would want to perform the integral in the first place, which is very important for students who will use calculus in other courses. We argue strongly here that infinitesimals– both chunky ones and those at the zero limit – should exist side-by-side in calculus courses offered to students who intend to pursue their studies in the physical sciences and engineering.

## 4.3 Symbolizing in the FTC

In a study of calculus students' understanding while problem-solving with definite integrals that involve physical quantities (velocity, force, energy and pressure), Sealey (2014) reports that "conceptualizing the product of $f(x)$ and $\Delta x$ proves to be the most complex part of the problem-solving process, despite the simplicity of the mathematical operations required in this step." Students struggled to understand how to form the product of two quantities, such as velocity and time, pressure and area, or force and distance, and how this product contributes to the overall calculation as an accumulation. We suspect that students' difficulties may stem from them not expecting the product to contribute to a quantity that accumulates (Thompson, 1994; Thompson and Silverman, 2008).

Sealey also notes that students did not struggle with the concept of a sum of elements going to infinity. Instead, it was the limit of the infinitesimal approaching zero that was problematic, and not the notion of limits writ large. Sealey included an orienting layer to her Riemann Integral Framework, which involves students making quantitative sense of the variables and quantities given in a problem before, and while, engaging in calculations. She found that students often revisited this layer throughout the problem-solving process, contributing to the growing body of evidence that the blending of physics quantity and the calculus was continual (Czocher, 2016; Schermerhorn and Thompson, 2023; Zimmerman et al., 2025).

Von Korff and Rebello (2012) emphasize the time it takes, and the significance of, symbolizing in the context of the FTC. In their experiment they found it necessary to provide direct instruction more than once to help the student understand the meaning of the symbols, but only at points when the student was ready to make meaning of those symbols. When the calculus symbolizing is combined with the many different letters used to represent both scalar and vector physical quantities, the representations require significant decoding, as exemplified in §3 with Gauss's law.

# 5 An FTC physics-framing for calculus instruction

A first step toward making the connections between calculus and physics explicit to students is to incorporate physical quantities into calculus instruction. We frame foundational physics quantities that students are already familiar with from prior instruction through their connections to calculus (see Table 2). These quantities can be framed through the FTC as change, rates of change and accumulations. A focus on quantities relies on Riemann sums of bits of accumulation made by a rate times a change as a sensemaking device for why we do calculus, not just how we do it.

In conjunction with modifications in calculus instruction, we envision careful, and mathematically correct, discussions in physics around select physical quantities, highlighting how calculus reasoning facilitates thinking about quantities as they change over a given interval, and the physical implications of that change. There is little time devoted in a typical physics course to helping students make these kinds of connections, and there should be.

We've presented findings from physics and mathematics education research suggesting that making symbolizing an explicit part of instruction will help students to better understand the quantities and operations they represent. In Table 3, we provide a structure so that instructors from both disciplines can draw attention to the symbolizing associated with ACRA, providing students adequate opportunity to fully comprehend sigma notation and indices, as well as $\Delta$ and $d$. We emphasize that this structure can guide instruction and discussion that may help address the challenges discussed in §4. In addition to symbols, we've included language and explicit reasoning around the quantification of change, rate and accumulation, and their representations in calculus and physics, that can be particularly valuable to students taking both courses. Table 3 extracts salient features of Table 2, and generalizes them such that the reasoning could be recognized across the many other contexts students will encounter in their subsequent coursework. In the remainder of this section we provide more detail for the structure of Table 3.

## 5.1 Change as physical quantity

One of the beauties of careful mathematical formalism is its generalizabilty. Unless the variables are quantities, there is no particular preference for which variable is the independent variable, designated by $x$, and which is the dependent variable, designated $y$. The RoC is quantified simply as the change in $y$ divided by the change in $x$ – which has the quantitative meaning that the change in $y$ is some number of times as large as the change in $x$.

The physical world adds a layer of constraint to the calculus it uses in that the *models must be testable*. That testability involves visualizing an experiment in which you manipulate the quantity represented by independent variable by changing its value, and predict the value of the dependent variable based on the function that relates them, and then measure the dependent variable to test the model. In addition, some quantities (e.g. position, time) are much easier to measure than others (e.g. energies). Nearly all of the quantitative functions students encounter in their first physics courses are functions either of position or of time.

**Table 3** FTC symbols and quantities common across calculus and physics

|  | operators | quantities | language | examples |
|---|---|---|---|---|
| change, interval | $d$, $\Delta$ | $dy$, $\Delta x$ | dep. variable change | impulse as change of momentum, |
|  |  |  | indep. variable interval | displacement as change of position |
| rate of change (RoC) | $\frac{\Delta}{\Delta x}$ | $\frac{\Delta y}{\Delta x}$ | ratio of change to interval of change | acc. as the time rate of velocity, |
|  | $\frac{d}{dx}$ | $\frac{dy}{dx}$ |  | force as the time rate of momentum |
| accumulation | $\sum$ | $\sum_i (\frac{dy}{dx})_i dx_i$ | sum of many small pieces | work, impulse |
|  | $\int_a^b$ | $\int_a^b f'(x) dx$ |  |  |

In the context of actual measurement, the scientist chooses an interval size ($\Delta x$) and measures the change in the dependent quantity ($\Delta y$) over that interval. While both ($\Delta y$) and ($\Delta x$) are considered "change" in mathematics (and reflected in Table 1), they are very different kinds of change. In the context of measurement one is manipulated and the other is a response, even though they covary. This relationship in experimentation is not unique to physics, so is generalizable across other science and engineering contexts.

We suggest referring to the change in the independent variable, ($\Delta x$), as an *interval* of change, emphasizing that the resulting change in the dependent variable, ($\Delta y$), depends on this interval (See Table 3). This framing aligns more naturally with experimental practices and broader reasoning in physics than a simplified input-output model might imply, and may help support more productive blending of mathematical and physical thinking. We propose a structure that focuses on *change* in the dependent variable and an *interval* of the independent variable. This framing is particularly useful in the context of physical quantity measurement, as described above.

Regarding these quantities in the context of the ACRA framework for conceptual learning of the FTC in calculus, we'd like to advocate for the inclusion of *interval* as a fundamental mathematical quantity as well (compare Table 1 to Table 3). It can serve both as a bridge between STEM models and prior calculus instruction, as well as giving more meaning in mathematics courses to the concepts of independent and dependent variables – supplementing their symbolic representations, positions on the axes of a graph, and framing as input-output. We extend the ACRA acronym to be ACRIA, to emphasize the importance of this distinction in experimental science.

We note that a common challenge in learning physics is that *change* is a quantity that is *different* from the quantity itself. The change in the energy or momentum of a system, or the change in the velocity of an object are commonly conflated with the quantities themselves (Rosenquist and McDermott, 1987). Several changes are so important in physics that they are given their own name – displacement as a change in position, impulse as a change in momentum, work as a change in system energy. We believe that early and frequent use of quantity in mathematics can help students recognize amount and change as different additive structures.

## 5.2 Rate as physical quantity

A RoC, seen in mathematics as the change in $y$ over the change in $x$, can be thought of as a change in $y$ over an interval $\Delta x$ – the change in $y$ that occurs as the independent quantity's value varies through an interval. This difference goes beyond semantics; it is more generalized than the everyday notion of time-rates, and it is how a physicist envisions rate. The concept of unit rate, "a change in the numerator for every unit of change in the denominator" is helpful here, especially when we apply proportional reasoning to accommodate non-unit changes in the denominator and when we apply smooth continuous variational reasoning to $dx$.

In some cases, the models generalize to physically meaningful situations in the ways mathematics allows, with no attribution as to why one quantity or another changes. One could think of a bidirectional causality – a small time interval for an object in motion will result in a small change in position. Or, conversely, a small displacement implies that there must have been a change in time. Position and its RoC, velocity, and its RoC, acceleration are names given to rates that help quantify motion, and carry no information about why the quantities change.

The conservation laws are different from the kinematics quantities, even though they are structurally identical mathematically. They all represent models in which the measured change of the independent variable is *physically caused* by the RoC in the integral. For the conservation laws, the change of momentum, and the change in the potential energy are due to the force exerted over an interval – the force is the time RoC of the momentum, and the position RoC of the potential energy. The meaning that these quantities carry are central to the conservation laws of physics. By contrast, acceleration is not a cause of a velocity change, just a time rate at which it happens. Unlike the kinematic quantities, manipulating energy or momentum intervals and finding the corresponding position or time change makes scant physical sense.

We propose a structure that focuses on RoC as a ratio of a change in the dependent variable in relation to an interval of the independent variable, as described above. It is important to emphasize that rate as quantity, is different from the quantity itself. In addition to the conflating a quantity and its RoC in the context of integration as described in §2, we note that it is also common for introductory physics students to conflate a quantity, its change and its RoC. Some examples of these distinctions are position, displacement and velocity as the time rate-of-change (RoC) of position; velocity, change-in-velocity, and acceleration as the time RoC of velocity; mass and density as the volume RoC of the mass, absolute pressure, pressure change and the volume RoC of the pressure, and so many more.

### 5.3 Accumulation as physical quantity

Visualizing the integral as summing up small bits is valuable in physics. Accumulations connect graphical representations, which are foundational to expert reasoning, with quantities that carry physical meaning (Zimmerman et al., 2025). The accumulation of the product of changing rates over short time intervals is at the foundation of kinematics. Similarly, the area under a force-position graph can be taken to represent an energy change when $\frac{dU}{dx}$ is taken to be a RoC of system mechanical energy with respect to position. The area under the a force-time graph can be taken to represent a momentum change when $\frac{dp}{dt}$ is taken as a RoC of momentum with respect to time. The interplay between these quantities is at the heart of the conservation laws of physics.

### 5.4 Physical quantities include units

We emphasize that amount, change, RoC, interval, and accumulation are all quantities with associated units, and that working meaningfully with these quantities requires attending to their units–both for students and for experts. In most calculus textbooks, kinematic variables are introduced with units, but these are often quickly abandoned once calculations begin. This practice is pedagogically flawed, especially for students studying or planning to study physics. Units are not auxiliary; they are integral to the very definition of a quantity and are essential for expert reasoning in applied contexts. Including units consistently throughout the calculation process is not merely helpful, it is crucial for developing a conceptual understanding of what calculus is doing. Take, for example, the equation $W = \int F dx$. A common student difficulty in physics is conflating force and work as similar or interchangeable quantities (Lindsey et al., 2009). If calculus instruction systematically includes units, students can begin to see that they are summing products of two distinct quantities–force (N) and displacement (m)–and that, through dimensional reasoning, work and force must be fundamentally different. Even though the precise distinction between them will be developed in physics, the epistemic framing, that units matter and help distinguish one quantity from another, lays a critical foundation for deeper learning.

We strongly urge instructors to use physics quantities – velocity, acceleration, force, position, displacement, energy change, time, time interval, and momentum change– and explicitly include units when teaching with foundational quantities in calculus courses. Supporting students in identifying the mathematical role of each quantity, whether as a change, a rate, an interval, or an accumulation, not only deepens their understanding of core calculus concepts but also prepares them to engage more productively with the scientific ideas these quantities represent.

# 6 Implications for research and development

## 6.1 Future directions in instruction and research

In §5 we present a quantities-based framing in Table 3 that can inform instruction of both integration and differentiation. It is often assumed that the "chunkiness" of the physical world is a concern beyond the scope of a first-year calculus course. However, we argue that engaging with this idea, particularly through the lens of infinitesimals, is not only appropriate but pedagogically valuable. One compelling rationale is that a significant portion of time in calculus courses is currently spent on manipulating compound functions that have little or no relevance outside of pure mathematics. In contrast, most functions that appear in real-world applications, particularly in introductory physics, are mathematically simple.

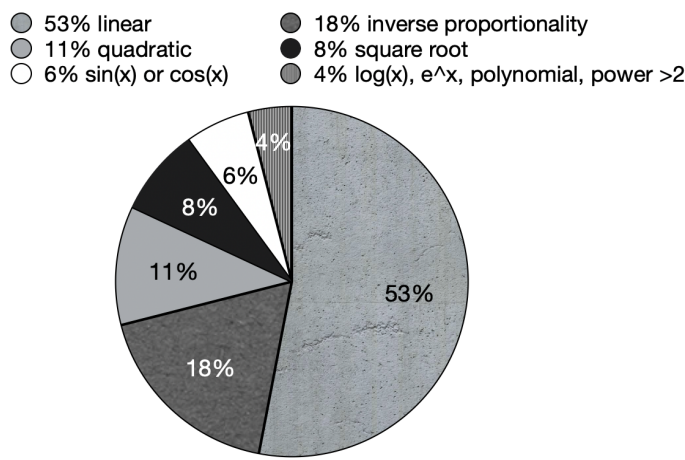

**Fig. 3** Distribution of functions in a typical introductory physics textbook (Elert, 2023)

A recent survey of a standard calculus-based physics textbook supports this view (see Fig.3). The overwhelming majority of models involve only linear, inverse proportionality, or quadratic functions, with over 80% of all functions falling into these basic categories (White Brahmia, 2023). This suggests that the emphasis on more complex, composite functions in introductory calculus may not reflect the kinds of reasoning students need when applying mathematics in scientific contexts. By reducing the time devoted to teaching mathematically sophisticated, but contextually rare, functions educators could instead foreground the conceptual development of the tools described in Table 3, including the symbols students encounter frequently but often struggle to interpret meaningfully.

Concerns may arise that introducing a "chunky" or physically grounded interpretation of $dx$ is inconsistent with the goals of a calculus course, which traditionally relies on abstract, limit-based definitions. However, recent research by McCarty and Sealey (2024) reveals a surprising degree of variation among expert mathematicians in how they conceptualize differentials. Notably, a significant subset reasoned about $dx$ in ways consistent with the physics perspective, that is, as a small but finite interval that can carry physical meaning. These findings suggest that there is room, even among experts, for multiple, context-dependent interpretations of differentials.

We propose that fostering this kind of flexible reasoning in students, especially the ability to invoke infinitesimals in a selective manner when appropriate to the context, should be considered a desirable learning outcome in calculus education. Rather than treating the infinitesimal as a relic of pre-rigorous calculus, we can embrace it as a bridge between the abstract mathematical world and the tangible physical one. Doing so not only aligns calculus instruction more closely with its applications but also equips students with a more versatile conceptual toolkit.

We hasten to add that our long list of proposals entails a number of research agendas regarding obstacles to implementing them in calculus or physics instruction and obstacles students could encounter in forming new meanings and ways of thinking. Among them are:

- In §2 we proposed that thinking with the FTC rests upon dispositions that students should develop in middle school and high school mathematics. Schools in the U.S. are particularly poor at developing these ways of thinking, among both high school students and teachers (Frank and Thompson, 2021; Thompson and Harel, 2021; Yoon and Thompson, 2020; Byerley and Thompson, 2017; Thompson and Carlson, 2017b; Yoon et al., 2015). We call for international studies to examine ways various countries support (or not) students' learning in middle and high school that is propaedeutic for later learning in a calculus that emphasizes the FTC.
- To what extent is the nature and content of students' understandings of various quantities a deciding factor in their recognition of situations as involving the FTC? For example, to what extent is the way they envision the way a quantity varies conducive or obstructive to envision it accumulating? In what ways must they understand quantities' relationships in a situation before it occurs to them there is a RoC between them? Do answers to these questions differ for different quantities?
- What types of support do mathematics or physics instructors need to highlight the FTC in ways we have suggested? Studies from Carlson's Pathways to Calculus project suggest it is possible for instructors to adapt, but it involves a great effort for them to rethink the ideas they teach and to redirect their thinking to what students understand from instruction (Carlson et al., 2023, 2024).

In what follows we consider both research yet to be explored, and research-validated materials that can help interested instructors and researchers move the work described in this paper forward.

## 6.2 Approximating with Riemann sums

In a Riemann sum approach, the central question becomes: *What counts as an acceptable approximation, and how can you tell when you have one?* In the study by Roundy et al. (2015), this question is explored in the context of disciplinary differences. How do engineers and physicists determine that an approximation is sufficiently accurate, and why might mathematicians reject that same approximation as inadequate? In §4, we describe research into how students reason with Riemann sums in physics contexts.

Despite the importance of this topic, there is a noticeable lack of research; studies in this area are scarce, leaving many open questions about how students learn to use Riemann sums with quantities, and what constitutes effective instructional practice in a calculus course. We view this as a fertile area for interdisciplinary collaboration and research. To begin addressing this gap, we describe a set of research-validated instructional materials designed to help seed what we hope will become a vibrant area of development.

In the development of their textbook, Thompson, Ashbrook, and Milner approached accumulation using Cauchy's notion of convergence rather than the more traditional Weierstrassian notion of limit (Thompson et al., 2019). In this framework, an integral approximation is considered "essentially equal to" the exact value when successive terms in the approximation differ negligibly, according to locally defined tolerance levels. This reframing aligns well with how approximations are used in applied contexts. Complementing this, Jones (2015) examined students' understanding of definite integrals through three interpretive lenses: area, antiderivative, and accumulation. He found that students who understood integrals in terms of accumulation were significantly more successful in making sense of physics and engineering applications than those who relied on area or antiderivative meanings. Similarly, Oehrtman (2008) developed an "approximation structure" that emphasizes limits in terms of the process of refining approximations. For example, students learn to estimate instantaneous rates of change by calculating average rates over smaller intervals and observing how these approximations improve, thus building toward a conceptual understanding of the derivative and the integral as limits. These materials are available as part of the Clear Calculus lab activity set developed by Oehrtman and colleagues at Oklahoma State University (Oehrtman, 2012), and in the free online textbook authored by Thompson, Ashbrook, and Milner (Thompson et al., 2019).

We envision instructors not only using these resources but also adapting and expanding them to meet the specific needs of their students. These examples demonstrate the potential for enriching the standard calculus syllabus with applications that emphasize approximation, accumulation, and real-world reasoning – key ideas at the intersection of mathematics, physics, and engineering.

## 6.3 Quantifying physics and conceptualizing mathematical operations

Integrating physics quantities into calculus instruction remains an underexplored area of research, with most existing studies focusing narrowly on kinematics. However, substantial questions remain unanswered: *What empirical evidence supports the cognitive benefits of engaging students in quantification within a calculus course?* and *How does the inclusion of diverse physical quantities align with, or potentially complicate, instructors' learning objectives in calculus?*

We are not suggesting that calculus instructors introduce new physics content into their courses. Rather, we recommend leveraging students' prior exposure to familiar physical quantities to support mathematical sensemaking. Instruction should emphasize the mathematical construction of quantities, such as momentum, velocity, or pressure, and the reasoning that justifies their quantification. For example: Why is momentum quantified as mass times velocity? Why is speed expressed as distance divided by time?

These kinds of questions invite students to engage in proportional reasoning, which is an essential cognitive resource for interpreting rates and accumulations. Consider momentum. Understanding it as the product of mass and velocity stems from the idea that both mass and speed contribute to an object's resistance to stopping. Doubling either quantity doubles the momentum, reflecting two simultaneous proportional relationships. Furthermore, because stopping an object requires opposing its motion, directionality becomes relevant, making it clear that momentum must be a vector quantity.

In addition to deepening students' understanding of individual quantities, instruction should help students explore the relationships among them. Questions such as, How is kinetic energy related to work? How is force related to momentum? or How is acceleration related to velocity? can guide students toward recognizing when and why calculus concepts like rate of change or accumulation are applicable. These relationships need not be the focus of entire lessons; even brief, targeted instructional moments can enhance students' capacity to interpret and model applied contexts more effectively.

While engaging students with physics-based ratio and product quantities such as energy or work may seem challenging in the traditional structure of a mathematics course, where such concepts are often treated as outside the discipline's scope, we argue that doing so is not only feasible but beneficial. To support this approach, we describe a set of existing instructional materials that can serve as productive starting points for instructors and researchers. These materials, we hope, will also serve to catalyze further curriculum development and empirical investigation within this important interdisciplinary space.

Physics Invention Tasks (PITs) are designed to engage students in authentic quantification by inventing meaningful quantities–typically ratios or products–to characterize physical systems (White Brahmia et al., 2024). Using data from carefully crafted contrasting cases, students identify invariants before formal instruction, supporting deeper conceptual understanding. PITs ramp from everyday contexts to core physics ideas such as velocity, acceleration, work, and momentum, and have been successfully field-tested at both pre-college and college levels.

PITs are grounded in the Inventing with Contrasting Cases (ICC) framework developed by Schwartz and colleagues, which promotes preparation for future learning by giving students productive opportunities to structure problems and recognize key patterns (Schwartz and Martin, 2004; Schwartz et al., 2011).

For example, in a "clown crowdedness" task, students invent a measure to describe how packed a bus is by comparing cases with varying numbers of clowns and bus sizes (Schwartz et al., 2011). Success requires coordinating *both* variables in a single quantity– leading to a RoC like density. In another study, Schwartz and Martin (2004) asked students to invent a measure for statistical spread using data sets with identical means but differing variability. Students' invented indices often resembled standard deviation, and this generative work improved later understanding and transfer.

These examples illustrate how invention tasks help students attend to structure and meaning, which is essential for ACRIA. PITs extend this approach to physics and calculus, helping students build conceptual foundations for key STEM ideas through invention, pattern recognition, and principled reasoning.

RoC, change, sum and product quantities can, and should, be explored prior to taking calculus. The Precalculus: Pathways to Calculus curriculum includes a textbook, workbook, and a range of supplemental materials designed to support students in constructing foundational calculus ideas–many of which are especially relevant to physics (Carlson et al., 2020). The curriculum emphasizes the development of concepts such as constant and changing rates of change through the lens of covariational reasoning, a research-supported approach that helps students understand how quantities change in relation to one another (e.g., Carlson et al. (2002); Thompson (1994)). Rather than prioritizing broad exposure to a wide array of function types, the materials focus on building a deep understanding of core functions through multiple representations–symbolic, graphical, numerical, and contextual–and through meaningful applications. Notably, the curriculum includes vector quantities, as well as sequences and series as tools for approximation, aligning with the kinds of reasoning required in physics contexts. By embedding mathematical ideas in authentic modeling situations and treating student knowledge construction as central, rather than an afterthought, the Pathways curriculum reflects contemporary research in mathematics education that highlights the importance of active, contextualized learning for conceptual development.

Lastly, these recommendations raise a broader question about the role of emerging technologies in enhancing learning. Platforms like Desmos (2011) and PhET Interactive Simulations (2002) offer dynamic alternatives to static graphs of flowing quantities, helping students visualize covarying change and relationships between physical quantities. While static representations have instructional value, they often fall short in conveying continuous change. Interactive tools can make abstract concepts–such as accumulation and rate of change–more concrete and accessible, and we encourage their use in instruction *in thoughtful ways*.

To grasp the idea of summing many small changes to approximate a total change, or to understand zero-limit rates of change, students must engage actively in *constructing* these concepts. Supporting that engagement is not straightforward. Technological tools have a learning curve, and are not simply a way to "see" abstract quantities. Students should learn not just how to operate the tool, but how to interpret what they are seeing; the technology must support their reasoning development. This long-standing challenge in STEM education highlights the importance of thoughtful instructional design.

# 7 Conclusion

Since the time of Newton, the disciplines of calculus and physics hae been deeply intertwined, yet their current instruction often misses opportunities for synergy. In our work, we demonstrate that the act of quantifying in physics–which frequently involves reasoning about change, rates, products, and sums–presents a substantial cognitive challenge. This challenge can result in cognitive overload when students hold unproductive or fragmented meanings for the mathematics used to construct symbolic representations. While procedural fluency in calculus is important, physics students often fail to see meaningful connections between the calculus they have learned and the physical phenomena they are asked to understand.

To address this disconnect, we propose a quantities-based framing of the Fundamental Theorem of Calculus (FTC) for both calculus and physics instruction. We argue that instructional time in calculus should be strategically reallocated to deepen students' understanding of integrals and derivatives as they relate to quantities–specifically, those that emerge from viewing the world through the lenses of rate of change (RoC) and accumulation.

One way to create space for this reconceptualization is to narrow the range of function types that students are expected to master procedurally. Another is to place greater emphasis on integrals as accumulations and derivatives as rates of change of accumulations. Our approach to the FTC is grounded in this idea: that RoC and accumulation are two sides of the same conceptual coin. The traditional treatment of the FTC in mathematics emphasizes a procedural link, that closed-form antiderivatives should be used to compute definite integrals. In contrast, our proposed approach highlights a quantitative connection: integrals represent accumulations resulting from quantities varying at (possibly non-constant) rates of change, and the quantities involved in a rate of change are themselves co-accumulating.

Further compounding the challenge, physics introduces additional mathematical complexity. Students must symbolize concepts and reconcile the use of vector quantities, despite encountering only scalar quantities in their prior mathematics education. Success in physics requires both procedural competence and deep quantitative understanding of rate, change, and accumulation, particularly as related through the FTC. Without a solid conceptual foundation in how these ideas connect, students struggle to learn these new mathematical layers in physics.

We argue that both disciplines would benefit from a shared instructional goal: for students to develop what Gray and Tall (1994) termed a proceptual understanding of calculus. This means that symbolic procedures and quantitative meanings become mutually evocative, thinking about one naturally invokes the other. In the context of definite integrals, such an understanding would allow students to fluidly move between symbolic expressions and their underlying quantitative interpretations. In this paper we presented an FTC physics-framing for calculus instruction in Tables 2 and 3 that can help calculus instructors find a path forward to fostering improved cross-disciplinary learning.

Fostering this kind of dual understanding promises to enrich student learning in both calculus and physics. By framing the FTC in terms of quantities, and by emphasizing the reciprocal relationship between rate and accumulation, we can support students in developing a more integrated and transferable understanding of change across disciplinary boundaries.

## Declarations

- On behalf of all authors, the corresponding author states that there is no conflict of interest.